\documentclass[traditabstract]{aa} 
\usepackage{graphicx}
\usepackage{txfonts}

\usepackage{natbib} 
\bibpunct{(}{)}{;}{a}{}{,} 
\usepackage{url}
\usepackage{multirow,bigdelim}
\usepackage{rotating}

\usepackage{subfigure}

\def \th {\thinspace}
\newcommand{\qc}{\mbox{4C +74.26 \th}}

\def \kms {\hbox{km s$^{-1}$}}

\def \ergsc{\hbox{erg s$^{-1}$ cm$^{-2}$}}

\def \ergs{\hbox{erg s$^{-1}$}}

\def \msun {\hbox{${\rm M_\odot}$}}
\def \msunyr{\hbox{{$M_{\odot}$ yr$^{-1}$}}}

\newcommand\omegam{\hbox{{$\Omega_{\rm m}$}}}
\newcommand\omegalambda{\hbox{{$\Omega_{\Lambda}$}}}
\newcommand\kmsmpc{{\rm km s$^{-1}$ Mpc$^{-1}$}}
\newcommand\ho{\hbox{{$H_{0}$}}}
\def \nh {\hbox{ $N{\rm _H}$ }}
\def \colc {cm$^{-2}$}

\def \xic {erg cm s$^{-1}$}
\def \fwhm {{\em FWHM}}
\newcommand{ \lia} {Ly$ \rm{\,\sc{\alpha}}$}
\def \fek {Fe K$\alpha$}
\def \ebv {\hbox{ $E{\rm _{B-V}}$ }}

\newcommand{\xmm }{{\rm XMM-\it{Newton}}}
\newcommand{\chandra }{{\it Chandra\th \th}} 
\newcommand{\heg }{{\it Chandra-\rm{HETGS} \th}} 

\begin{document}

\title{The warm absorber in the radio-loud quasar \qc}
\subtitle{}

\author{L. Di Gesu \inst{1,2}
       \and E. Costantini \inst{1}
}

\institute{
SRON Netherlands Institute for Space Research, Sorbonnelaan 2, 3584 CA Utrecht, The Netherlands 
\and
Department of Astronomy, University of Geneva, 16 Ch. d' Ecogia, 1290, Versoix, Switzerland \email{Laura.DiGesu@unige.ch}}
\date{Received April 8, 2016; accepted June 13, 2016}
\abstract
{
Outflows of photoionized gas
are commonly
detected in the X-ray spectra
of Seyfert 1 galaxies. However,
the evidence for this phenomenon in
broad line radio galaxies, which are  analogous
to Seyfert 1 galaxies in the radio-loud regime, has so far
been scarce.
Here, we present the analysis of the X-ray absorption 
in the radio-loud quasar 4C +74.26.
With the aim of characterizing  the kinetic 
and the ionization conditions
of the absorbing material, 
we fitted jointly  the XMM-\textit{Newton} Reflection Grating
Spectrometer (RGS) and
the \textit{Chandra} High Energy Transmission Grating
Spectrometer (HETGS) spectra, 
which were taken 4 months apart. The
intrinsic continuum flux did not vary 
significantly during this time lapse.
The spectrum
shows the absorption signatures
(e.g., Fe-UTA, 
\ion{O}{vii}, 
and
\ion{Ne}{vii}--\ion{Ne}{x}) 
of a photoionized
gas outflow 
($\nh \sim 3.5 \times 10^{21}$ \colc,
$\log \xi \sim 2.6$,
$v_{\rm out}\sim 3600$ \kms)
located at the redshift
of source.
We estimate that
the gas is located outside the broad line region
but within the boundaries of the putative torus.
This ionized absorber is consistent 
with  the X-ray counterpart 
of a polar scattering outflow reported
in the optical band for this source.
The kinetic luminosity carried by
the outflow is insufficient
to produce a significant
feedback is this quasar.
Finally, we show that
the heavy soft X-ray
absorption that was noticed in the past
for this source arises mostly
in the Galactic ISM.} 
\titlerunning{}
\authorrunning{L. Di Gesu et al}
 \maketitle
%
%

\section{Introduction}
\label{intro5}
In the last fifteen years,
thanks to the advent of high-resolution X-ray spectrometers,
such as
the \xmm \th \th Reflection Grating
Spectrometer (RGS) or
the \chandra Low and High
Energy Transmission Grating
Spectrometers (LETGS and HETGS),
our knowledge of
the circumnuclear gaseous
environment of active galactic
nuclei (AGN) 
has advanced significantly.\\
It is now established that
roughly half 
of all local Seyfert galaxies
host a photoionized warm
absorber (WA) that produces
features detectable in
the X-ray and in the UV
band \citep{cre2003}.
These absorption lines
are usually blueshifted
with respect of the systemic
velocity, which indicates
a global outflow of the
absorbing gas.
Spectroscopical observations
allow  the
physical conditions
(kinematics and ionization)
of the gas to be characterized with high accuracy
\citep[see][for a review]{cos2010}.
In photoionization equilibrium, 
the ionization parameter 
$\xi=L_{\rm ion}/ n r^2$
(where $L_{\rm ion}$ is the
ionizing luminosity between 1 and 1000
Ryd, $n$ is the gas density, and $r$
is the distance from the ionizing source)
parameterizes the state 
of the gas. In the X-ray
band, there are many transitions, for example
from ionized
C, N, O, Ne, and Fe,
which allow  
an accurate solution
for $\xi$ to be determined.
From spectroscopical observables,
useful constraints can be put
on the gas location
\citep{blu2005};
these constraints   quantify
how much momentum is transferred
by the outflow
to the surrounding medium
\citep[e.g.,][]{cre2012}.\\
%
%
The studies of WA in
Seyfert galaxies
show that these outflows
span roughly four
orders of magnitude in ionization
($ \log \xi \sim 0 - 4$)
and reach velocities 
of a few thousand \kms
\citep{mck2007}.
They are often located
as far as the putative
torus \citep{blu2005}. 
Some outliers
may be located closer to the
nucleus, at the distance
of the accretion disk
or farther out in the galaxy
at $\sim$kpc distance from
the center \citep[][]{dig2013}.
In most of the cases,
the kinetic power of the
WA is found to be negligible
with respect to the AGN radiative
power \citep[e.g.,][]{ebr2016}. 
Thus, WA are not expected
to play a significant role
in a possible negative
AGN feedback \citep{sca2004,som2008,
hop2008,hop2010,kin2015}. \\
A different class
of photoionized winds
are the  ultrafast
outflows (UFO). These
may be present in 35\% of Seyfert galaxies \citep{tom2010} 
 and 
differ from classical
WA because of the higher outflow
velocity (v$\sim$0.1 c, where c is the speed
of light) and of the higher ionization 
($\log \xi \geq 3 $, \citealt{tom2011}).
Hence, because of the higher
energy and higher blueshift 
of their transitions 
(e.g.,
\ion{Fe}{xxv}--\ion{Fe}{xxvi}),
UFO  are detectable
only in lower resolution
CCD spectra.
These powerful winds
are believed to be a nuclear
phenomenon originating
from the accretion disk
\citep{tom2012, nar2015}.\\
%
%
The detection of 
photoionized features
in broad line radio galaxies
(BLRG),
which are  analogous
to Seyfert 1 galaxies in the  radio-loud regime, was expected
to be difficult because
of the presence of a
relativistic jet. 
The Doppler-boosted, 
non-thermal radiation 
of a jet located close
to the line of sight 
could actually mask the absorption features.
So far, the statistics
of known WA in BLRG relies
on a handful of cases,
of which only three
are WA detections in
a high-resolution
X-ray dataset.\\
Hints of 
photoionized absorption
were  noticed,
for instance,
in the ROSAT-PSPC spectrum
of 3C 351
\citep{fio1993}
and 3C 212 \citep{3c212}. 
Interestingly, these
two sources also display
WA features
 in the UV
\citep{math1994, yua2002}.
More recently, \citet{mol2015}
have reported the detection
of \ion{O}{vii}
and
of \ion{Fe}{xx} absorption
edges in the EPIC-pn spectrum
of IGR J14488-4008,
a giant radio-loud
galaxy discovered
by INTEGRAL.\\
The first case
of a WA in a BLRG
studied with a grating
spectrum
was a long
\heg spectrum 
of 3C 382 \citep{ree2009}.
The detection of this WA,
whose location is consistent
with the distance
of the narrow
line region (NLR), 
was promptly
confirmed by a subsequent
RGS observation
\citep{tor2010}.
A second case is 
the remarkable 
photoionized outflow in
3C 445.
In the \heg spectrum
of this source \citet{ree2010}
detected a low-ionization
outflow  moving at
a sub-relativistic velocity.
A deep Suzaku spectrum
also shows  indications
of blueshifted absorption
from highly ionized iron
\citep{bra2011}.
Both these spectra 
are consistent 
with a scenario where
our line of sight intercepts
an equatorial disk wind
located at $\sim$sub-pc
scale. 
The low-ionization 
absorber may consist
of sparse clumps
embedded in a
highly ionized wind
\citep{ree2010}.
In addition to these two cases, 
\citet{tor2012}
report a WA detection
in the RGS spectrum
of 3C 390.3.\\
Signatures 
of more highly ionized
UFO have also been
 detected
in the CCD spectra
of a handful of radio-loud
sources \citep{tom2014},
with a statistical
incidence comparable,
within the uncertainties,
to what is found for radio-quiet
Seyferts.\\ \\
In this paper we
present the analysis of the
X-ray grating spectrum --
obtained with the RGS
and the \chandra-HETGS --
of the BLRG
4C +74.26. 
This source is located at
a redshift of 0.104 
\citep{ril1988}.
In the optical, it
shows broad permitted
lines with a
a \fwhm \th of 10\,000 \kms\
for the H$_{\rm \beta}$
line \citep{win2010}.
Using this line width
a SMBH mass of 
$3 \times 10^{9}$ \msun\
\th is inferred.\\
Because of
its $\sim$1 Mpc 
projected linear size
\citep{ril1988},
this source is the
largest known
radio source associated 
with a quasar. Its
radio morphology
is typical for a
Fanaroff-Riley
type II source
(FRII),
although 
the 178 MHz
radio luminosity
is borderline
with the 
type I class
(FRI).
Observations
with the Very Large Array (VLA)
have revealed a one-sided 
jet which is at least 
4 kpc long \citep{ril1990}. 
The flux
limit for a counter-jet,
which could be set with a subsequent 
Very Long Baseline Interferometry
(VLBI) observation \citep{pea1992},
implies that the source axis
lies at $\la 49^{\circ}$ from 
our line of sight.\\
Evidence 
of a high-velocity outflow
in \qc was  found
in the optical spectropolarimetric
analysis performed in \citet{rob1999}.
These authors noticed that the broad
H$\alpha$ line appears redshifted in polarized light,
which can be explained if the scattering medium
producing the polarization is part 
of a polar outflow. \\
Since 1993, 
\qc has been targeted
by many X-ray observatories,
including 
ROSAT, 
ASCA,
\textit{Beppo}-SAX 
\xmm,
and \textit{Suzaku}.
In the \xmm \th \th
\citep{bal2005}
and in the Suzaku 
\citep{lar2008}
spectrum a 
broadened \fek \th emission line
has been clearly detected
at 6.4 keV.
Recently,
in  the Suzaku
\citep{gof2013}
and the \xmm \th \th
spectrum
\citep{tom2014}
additional absorption
features 
in the Fe-K band
have been  noticed.
These could be
due to a highly ionized
UFO, with a measured
outflow velocity
on the order of
$\sim0.1c$.\\
By studying the correlations
between the Suzaku light curves
in different bands, \citet{nod2013}
were able to extract 
the stable soft-excess component \citep{sin1985}
 that  may dominate
the continuum emission at
soft energies  (i.e.,  below 2.0 keV).
According to these
authors, the most likely origin for the
soft-excess in this source
is thermal Comptonization
of the disk photons in a warm plasma
\citep[see, e.g.,][]{nod2011, 
don2012, jin2012, pet2013, 
dig2014, giu2015, boi2016}.\\
It was  found, however,  that
the soft-excess  underlies a
heavy soft X-ray absorption.
For instance,
absorption from a substantial 
column density of gas
in excess at the Galactic column density
was seen earlier on 
in the ROSAT-PSPC \citep{bri1998},
ASCA \citep{bri1998,sam1999,ree2000},
and Beppo-SAX \citep{has2002} spectra.
In a more recent \xmm \th \th
observation,
\citet{bal2005a} detected
a column of cold absorption
greater than the Galactic value,
with an intrinsic column
of $\sim 1.9 \times 10^{21}$ \colc.
Moreover,
the broadband \xmm \th \th
spectrum shows evidence 
of a weak WA
intrinsic to the source.
The WA is highlighted 
by features identified 
as the \ion{O}{vii} and
\ion{O}{viii} absorption edges
\citep{bal2005}.\\
Motivated by these
indications of a complex
absorption in this source,
here we use the archival \xmm \th \th
Reflection Grating Spectrometer
(RGS) and \chandra
High Energy Trasmission Grating Spectrometer
(HETGS) spectra of 4C +74.26
to characterize for the first time
the kinematics and the ionization
condition of the X-ray absorbing
material.\\
In 
Sect. \ref{dataqc} we
describe our data
reduction
procedure.
Then  in
Sect. \ref{sedqc}
we build the
spectral energy distribution
(SED), and
in Sect. \ref{specqc}
we perform the spectral analysis.
Finally, in Sect. \ref{dsou} 
we discuss our results,
and in Sect. \ref{concqc} 
we state the conclusions.\\
The C-statistic \citep{cas1979} 
is used throughout the paper,
and errors are quoted at the 68$\%$ 
confidence level
($\Delta C=1.0$). In all the
spectral models presented,
we use the total Galactic 
hydrogen column density
from \citet[][$\nh=2.31 \times 10^{21}$ 
\colc]{wilh22013}. 
In our luminosity calculations
we use a cosmological redshift of z=0.104
and a flat cosmology with the following
parameters:
\ho=70 \kmsmpc, 
\omegam=0.3, and
\omegalambda=0.7.


\begin{table*}
\caption{XMM-Newton and \chandra observation log for 4C +74.26.}     
\label{obs5.tab}      
\centering                    
\begin{tabular}{lccccc}        
\hline\hline                 
Date & Instrument & Observation ID &
Net exposure \tablefootmark{a} & 
F$_{\rm 0.3-2.0 \, keV}$\tablefootmark{b} &
F$_{\rm 2.0-10.0 \, keV}$\tablefootmark{b} \\
& & & (ks) & \multicolumn{2}{c}{($10^{-11}$ \ergsc)} \\
\hline
2003 Oct 6 & HETGS & 4000 & 37 & 0.7 & 2.8 \\
2003 Oct 8 & HETGS & 5195 & 31 & 0.8 & 2.9 \\
2004 Feb 6 & RGS   & 0200910201 & 34 & 0.9 & 3.0  \\
\hline 

\end{tabular}
\tablefoot{
\tablefoottext{a}{Resulting exposure time after correction for background flares.}
\tablefoottext{b}{Observed flux in the quoted bands.}
}
\end{table*}


\begin{figure}[t]
    \includegraphics[angle=180,width=0.5\textwidth]{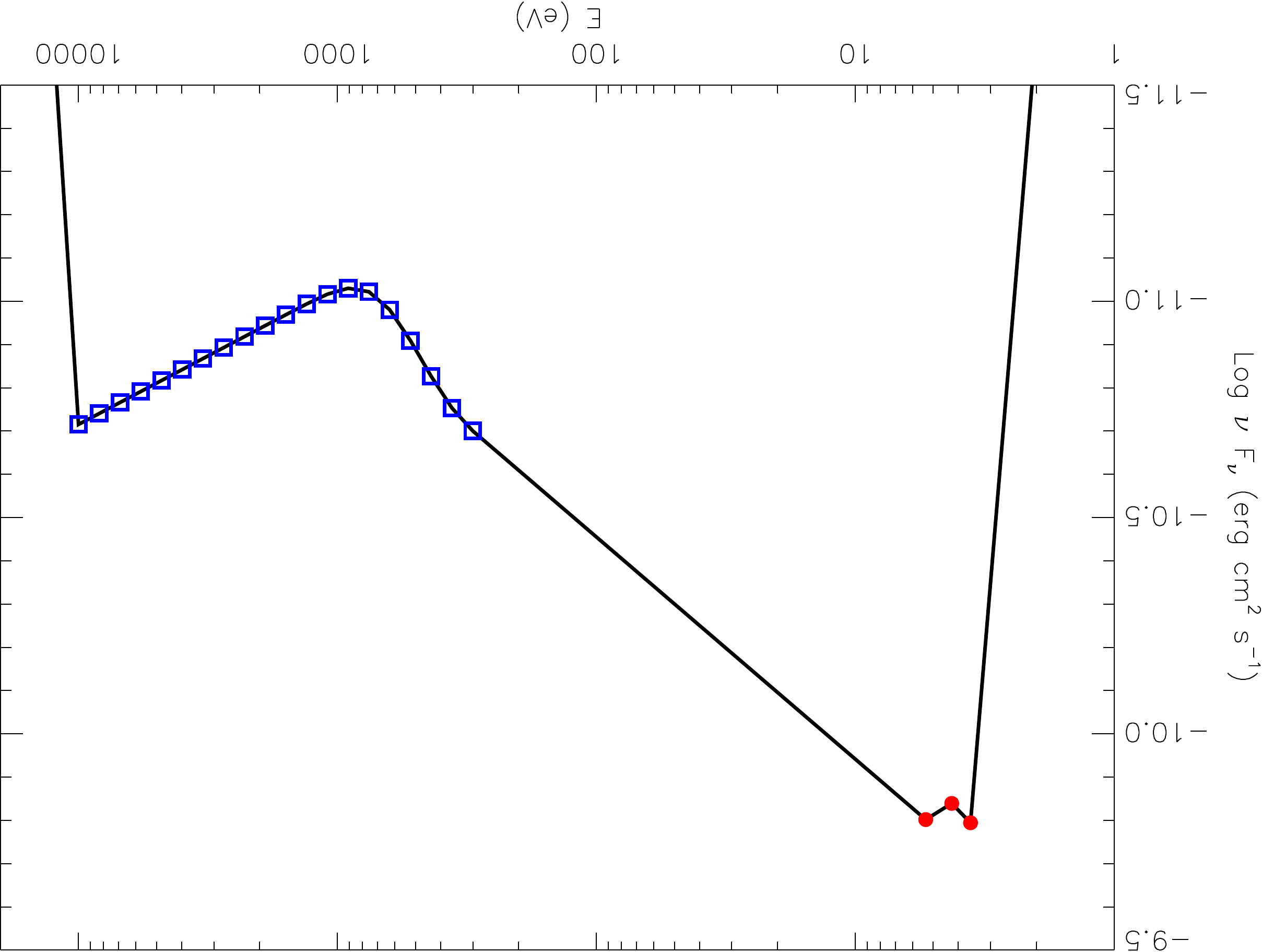}
  \caption{Spectral energy distributions for 4C +74.26.
   Filled circles: OM fluxes corrected for the Galactic extinction.
   Open squares: X-ray intrinsic continuum obtained from a phenomenological fit
   of the EPIC-pn spectrum.}
  \label{sedqc.fig}
\end{figure}


\section{Observations and data preparation}
\label{dataqc}
The radio-loud galaxy 
\qc was observed with \chandra and \xmm
\th \th in October 2003 and February 2004, respectively. 
Thus, the time separation between 
these X-ray observations is only 4 months.
In Table \ref{obs5.tab} we summarize the basic
information of each
observation. \\
\chandra observed \qc for $\sim$70 ks in total
using the HETGS in combination 
with the ACIS detector.
The total exposure time was 
split into two observations
that were taken two days apart. 
For both Obs-ID 4000 and 5195
we retrieved the Medium (MEG) 
and High Energy Grating (HEG) spectra 
and their respective response matrices
from the 
\textit{tgcat}\footnote{\url{http://tgcat.mit.edu/}} archive.
We further treated these spectral products
with the CIAO tools 
(version 4.6).
For each observation and for both HEG and MEG,
we combined the first positive and negative
spectral order using 
the CIAO script \textsf{add\_spectral\_orders}.
Hence, we fitted
jointly HEG and MEG 
(allowing a free intercalibration factor)
with a simple phenomenological
power law to check for variability between the
two observations. The fitted slopes
($\Gamma_{4000}=1.32 \pm 0.02$,
$\Gamma_{5195}=1.34 \pm 0.02$) 
 and normalizations
($Norm_{4000}=(9.9 \pm 0.1)
\times 10^{53} \, 
\rm ph \, s^{-1} \, keV^{-1}$,
$Norm_{5195}=(10.5 \pm 0.2)
\times 10^{53} \, 
\rm ph \, s^{-1} \, keV^{-1}$) 
were very consistent with each other
(see below for a physically motivated fit).
Therefore,
we were able to sum  the spectra of individual 
observations into a single spectrum 
to improve the signal-to-noise ratio.
We did this using the CIAO script
\textsf{add\_grating\_spectra}.\\
We reduced the raw \xmm \th \th
Observation Data Files (ODF),
available 
at the ESA archive\footnote{\url{http://xmm.esac.esa.int/xsa/}},
using the Science Analysis Software
(SAS, version 13) and the
HEASARC FTOOLS. 
We created calibrated 
EPIC-pn event files
selecting only the 
unflagged single events.
To check the time stability
of the background we used 
the light curve in 
the hard 10--12 keV
band, which is background
dominated. A high level of 
background light due to soft proton
contamination is evident
towards the end 
of the observation.
Thus, we cleaned the
event file using a time filter,
following the same procedure
explained in \citet{dig2013}.
For RGS-1 and RGS-2
we created calibrated 
event files and background
light curves taking
the background from CCD 9.
The RGS background
light curve was quiescent.
Next, for all
the instruments, we extracted
the source and background spectra
and we created the spectral
response matrices.\\
Finally, we extracted the source
count rate in all the available
OM filters, 
namely 
$U$ ($\lambda_{\rm eff}=3440$ \AA), 
$UVW2$ ($\lambda_{\rm eff}=2910$ \AA)
and $UVM2$ ($\lambda_{\rm eff}=2310$ \AA).
Using the interactive SAS tool
\textsf{omsource}, 
we computed the source count
in a circular region centered
on the source coordinates and
with a radius of 6 pixels.
For the background 
we used another circular 
region of 12 pixels, free
from other sources 
and instrumental 
contamination.
We converted the count rates to fluxes using
the standard conversion factors provided
in the SAS watchout web page\footnote
{\url{http://xmm.vilspa.esa.es/sas/7.1.0/watchout/Evergreen_tips_and_tricks/uvflux_old.shtml}}. 
Hence,
assuming $R_{\rm V}$=3.1,
we corrected all
the fluxes for
the Galactic reddening
\citep[\ebv=0.39, ][]{sch2011}.
For the correction,
we used the IDL routine
\textsf{ccm\_unred},
which dereddens a user-defined
vector of fluxes using
the Galactic
extinction curve
of \citet{car1989}.

\begin{figure*}[t]
\begin{minipage}[c]{1.0\textwidth}
 \includegraphics[width=0.5\textwidth]{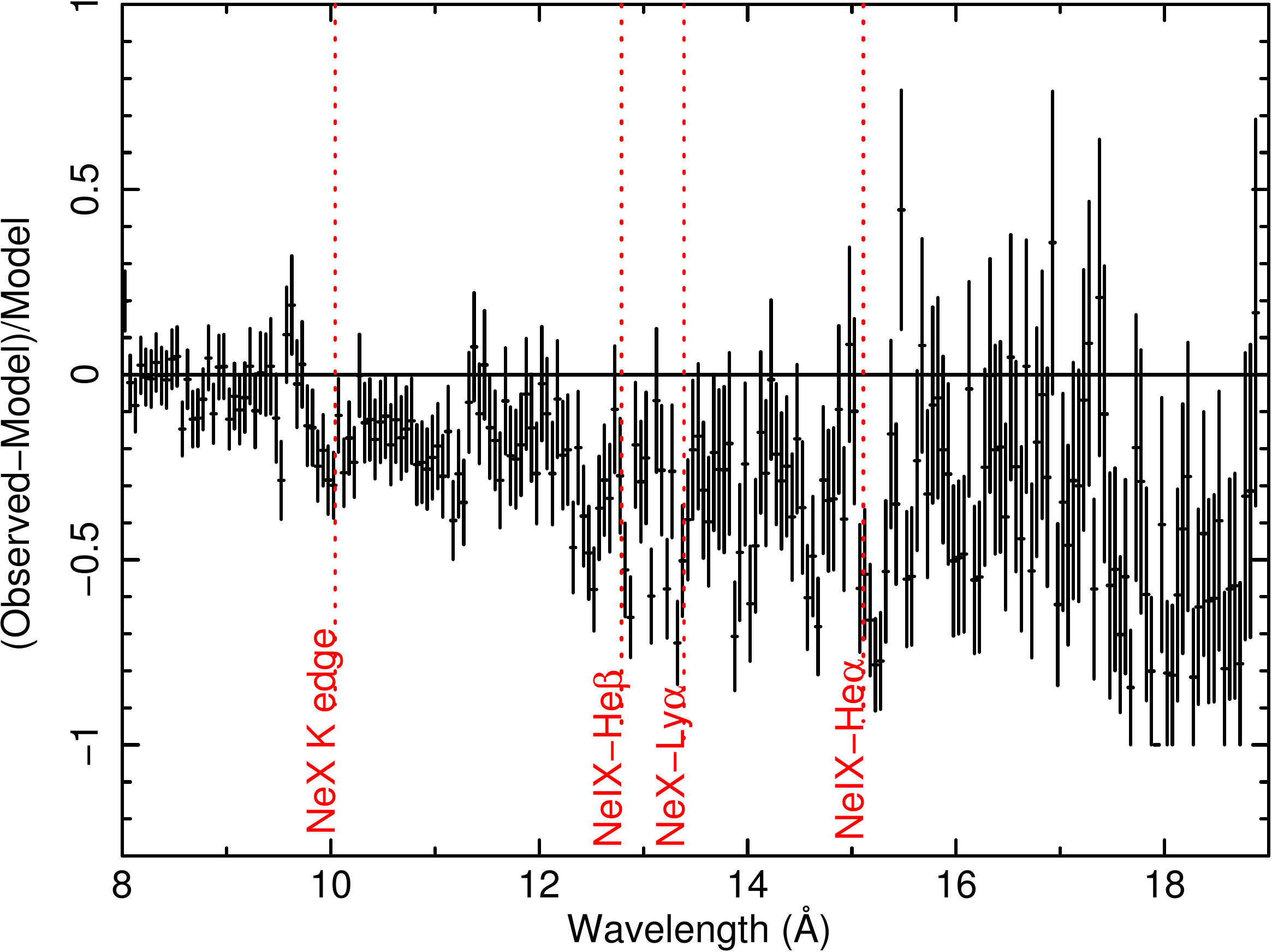}
 \hspace{0.3cm}
 \includegraphics[width=0.5\textwidth]{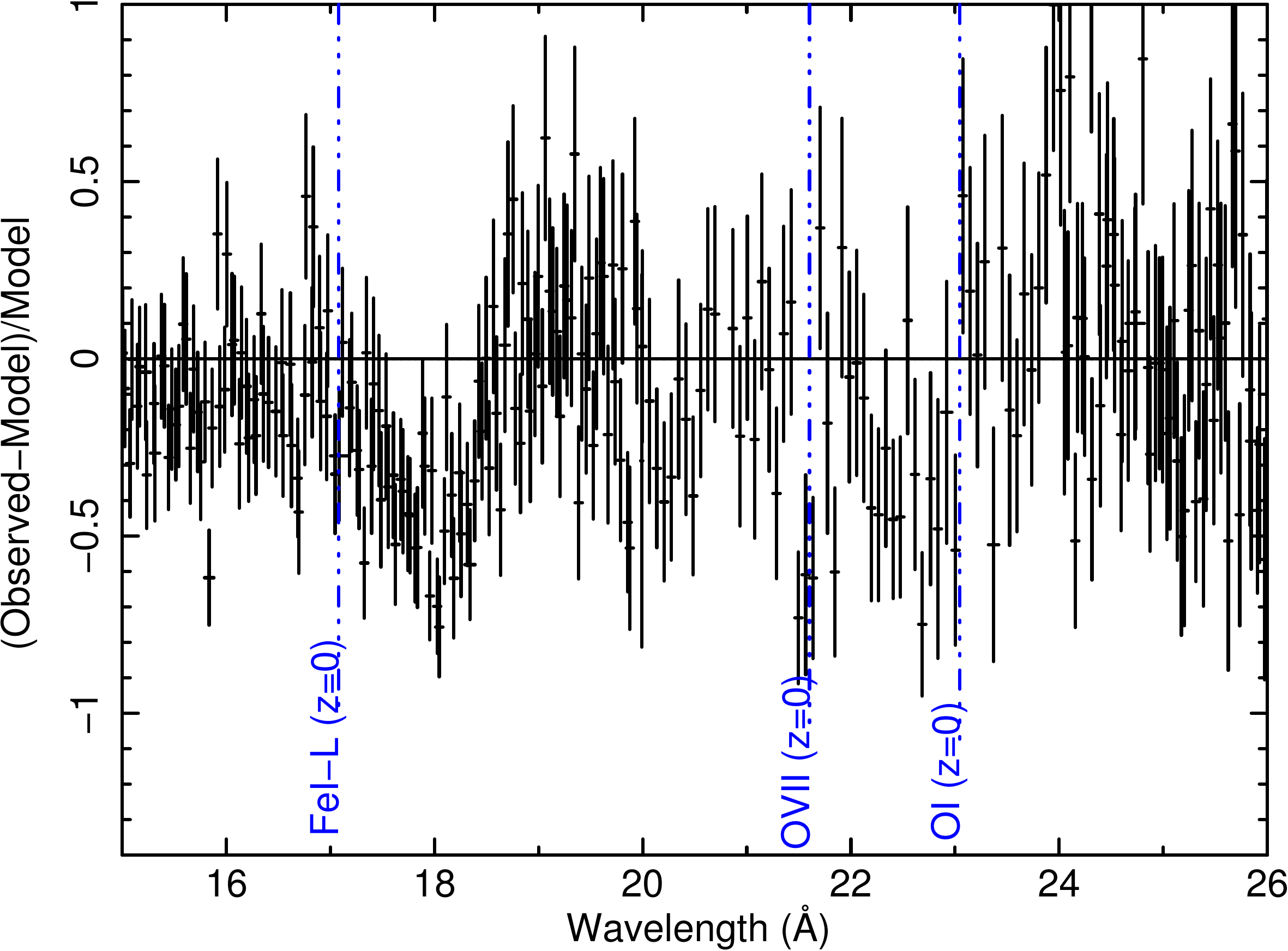}
\end{minipage}
  \caption{Relative residuals of the MEG (\textit{left panel}) 
   and the RGS (\textit{right panel}) spectrum 
   after a simple power-law fit. Vertical lines indicate
   the wavelengths of the Galactic (triple-dot-dashed line)
   and intrinsic (dashed line) candidate absorption lines and edges.
   The spectra have been rebinned for clarity.}
  \label{resqc.fig}
\end{figure*}



\section{ Spectral energy distribution}
\label{sedqc}
As the preliminary step of our analysis
we constructed the
spectral energy distribution (SED) of the
source.
This is needed
for the photoionization modeling of the
absorbers.\\
The OM fluxes
together with the EPIC-pn spectrum 
constrain the SED
from optical/UV up to X-ray energies.
We fitted the EPIC-pn data with
a phenomenological model including
a black-body 
($T_{\rm BB}=109 \, \rm keV$) at soft energies,
a power law ($\Gamma=1.7$) at hard energies,
and a broad \fek \th line (\fwhm=0.5 keV).
All these components are absorbed 
by the Galactic column density of 
\nh=2.31 $\times 10^{21}$ \colc\
(see the discussion below).
We adopted 
the unabsorbed phenomenological continuum
of this fit as the  X-ray SED.
Combining this 
X-ray continuum 
(Fig. \ref{sedqc.fig}, open squares)
with the OM fluxes corrected for the Galactic
extinction (Fig. \ref{sedqc.fig}, filled circles)
we obtained the SED shown in
Fig \ref{sedqc.fig}.
We cut off the SED at low
and high energy, respectively
at 0.01 Ryd and 100 keV.


\begin{figure*}[t]
\begin{minipage}[c]{1.0\textwidth}
 \includegraphics[width=0.5\textwidth]{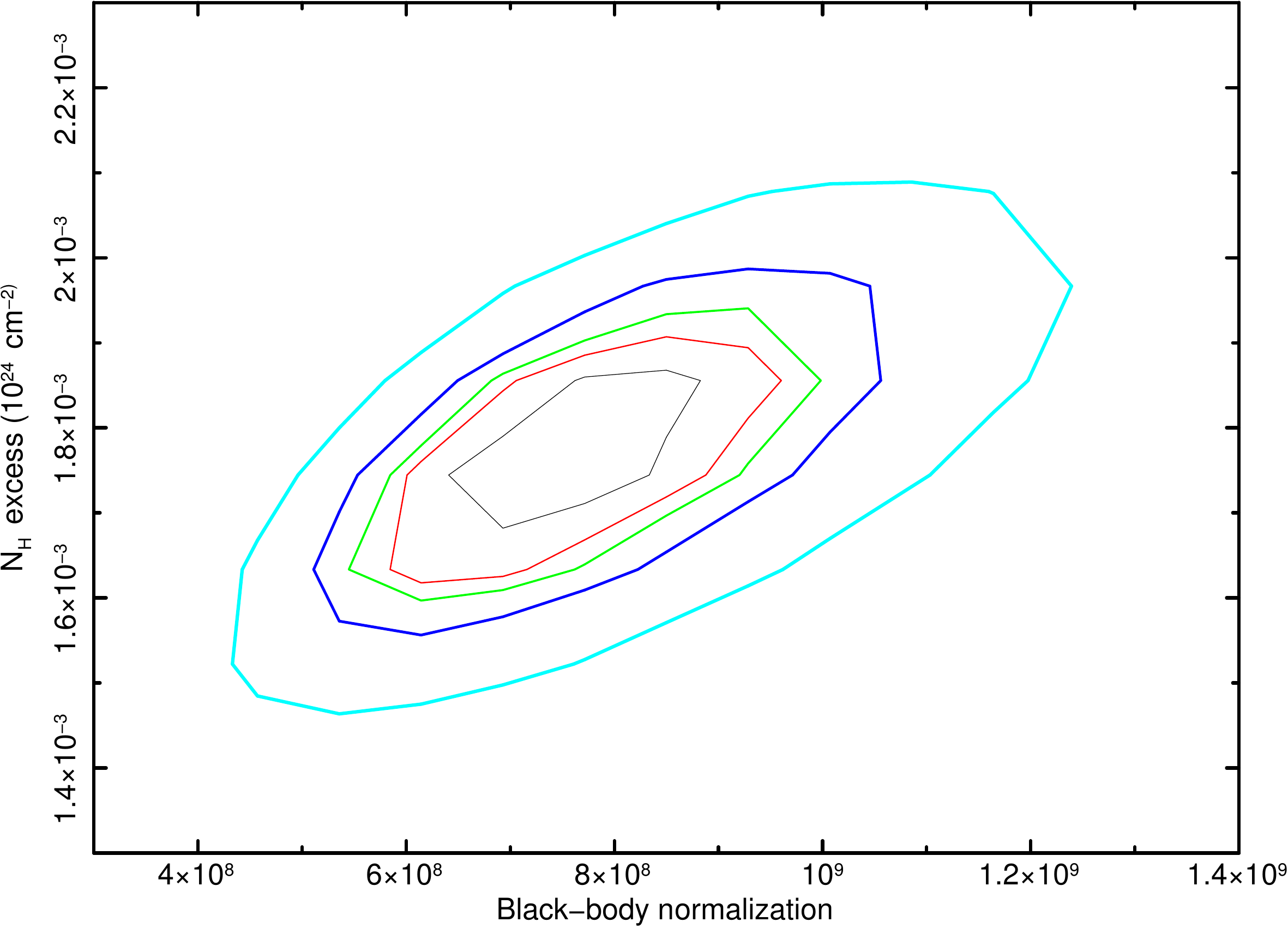}
 \hspace{0.3cm}
 \includegraphics[width=0.5\textwidth]{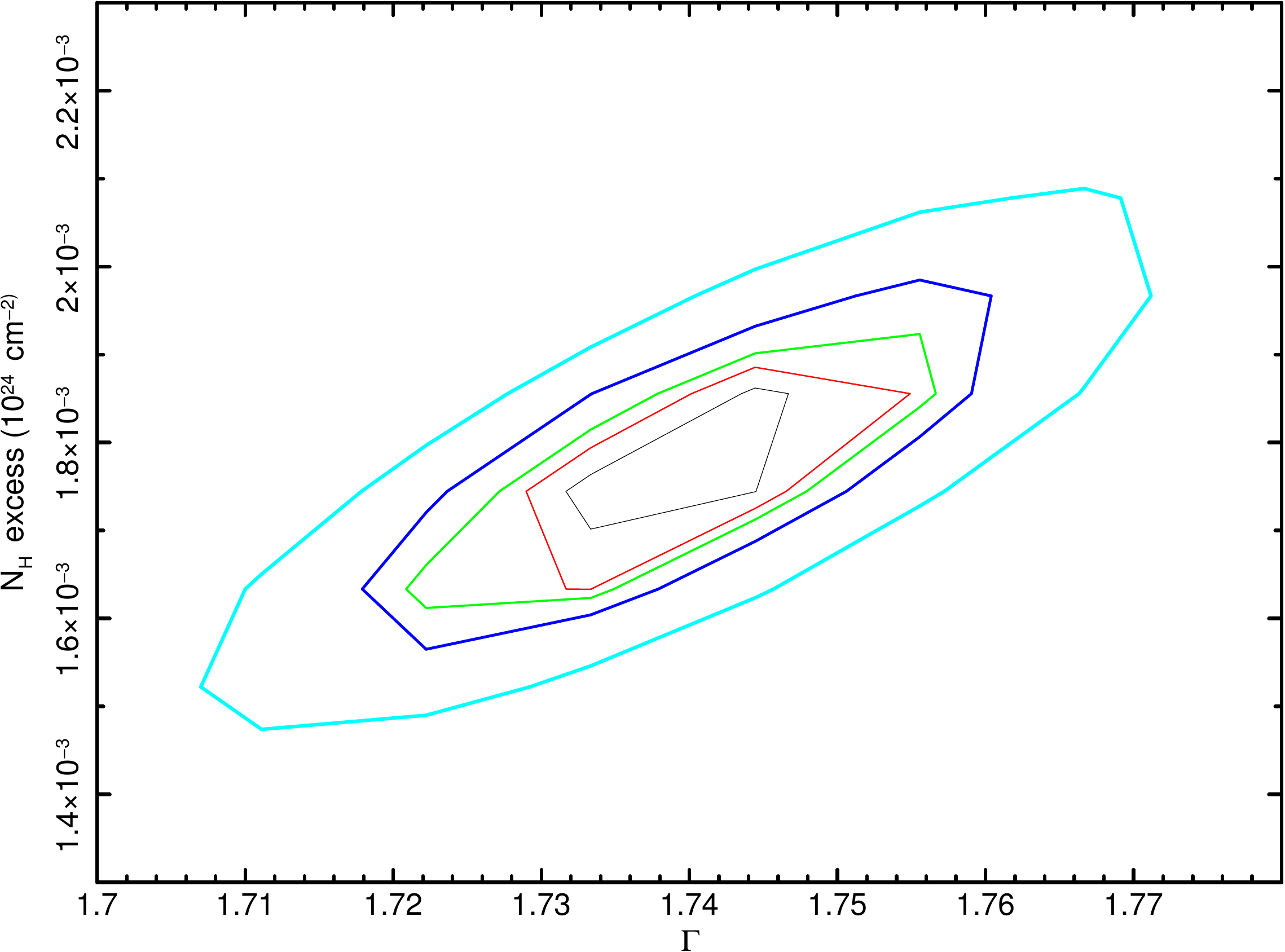}
\end{minipage}
  \caption{Confidence level contour plot for the
  excess of \nh at the Galactic value of \citet{kal2005} vs the normalization
  of the modified blackbody (\textit{left panel}) and the power-law
  slope (\textit{right panel}). The curves were obtained
   from a phenomenological fit of the EPIC-pn (Sect. \ref{sedqc}). 
   In each panel,
  the curves corresponding to a confidence level
  of 68\%, 90\%, 99\%, and 99.99\% are shown. }
  \label{contu.fig}
\end{figure*}


\section{Spectral analysis}
\label{specqc}

\subsection{ Galactic absorption}

The X-ray spectrum of 4C +74.26 
showed a heavy soft X-ray absorption
in excess at the Galactic column
measured by 21 cm surveys
(e.g., $\nh=1.16
\times 10^{21}$ \colc, \citealt{kal2005})
in all the historical records 
(Sect. \ref{intro5}).
We illustrate this by using
our phenomenological fit of the
EPIC-pn spectrum (Sect. \ref{sedqc}).
In Fig. \ref{contu.fig}, we show
the confidence contour (i.e., curves
of constant $\Delta$ C) of 
the \nh excess as a function
of the modified blackbody normalization (left panel)
and of power-law slope (right panel) .
An excess of \nh of at least $\sim 1.5 \times 10^{21}$
 \colc is observed in both figures
at a confidence level
of 99.99\%.\\
Part of this excess of absorption
can be due to the gas in our
Galaxy rather than to some absorber
intrinsic to the source.
The total foreground
X-ray absorption may be,
in some cases, significantly larger
than the value inferred using
the \nh value provided
by 21 cm surveys \citep{kal2005, dic1990}.
The difference 
may be ascribed to the presence
of hydrogen in molecular 
form (H$_{2}$) in the
Galactic ISM \citep{ara1999}, which  is indeed elusive 
to 21 cm measurements. 
We used the calibration
of \citet{wilh22013}
\footnote{\url{http://www.swift.ac.uk/analysis/nhtot/index.php}}  
to infer the equivalent
hydrogen column
density of the the molecular
hydrogen ($N_{\rm H_{2}}$)
along the line of sight 
of 4C +74.26.
We found $N_{\rm H_{2}}=
1.15 \times 10^{21}$ \colc.
Thus, the total Galactic hydrogen
column density absorbing
the X-ray spectrum is
$\nh+N_{\rm H_{2}} =2.31 \times 10^{21}$
\colc, consistent with the
total column density inferred
from the broadband X-ray
spectrum \citep{bal2005a}.
Applying the standard Galactic
\ebv/\nh ratio
\citep[$1.77 \times 10^{-22}$,][]{pre1995},
this hydrogen column density is also consistent
with the Galactic reddening
of \ebv=0.39
\citep{sch2011}. We
use this Galactic column
density value 
in all the spectral analyses performed
here. The remainder of the excess absorption 
is due to the photoionized gas of the outflow (see Sect. \ref{wa}).
In fact, as our model shows, 
there is no need for additional neutral gas 
when this outflow is taken into account.\\

\subsection{Preliminary  spectral residuals}
\label{resqc}
We performed the spectral
analysis of the RGS and the HEG
datasets using 
SPEX, version 3.0
\citep{kaa1996}.
We began by fitting
the RGS spectrum with a simple power-law continuum
absorbed by the Galactic hydrogen column density
and we inspected the relative residuals
(Fig. \ref{resqc.fig}, right panel).\\
The most prominent
features in the RGS
residuals
is a broad absorption trough
visible at $\sim 18$ \AA.  
We note that
the Fe-L edges from
the neutral absorber in our Galaxy 
cannot be responsible for this feature, 
as they would be expected
at $\sim$17.1 \AA.
Moving redward,
a narrow feature is clearly visible
located at 
the wavelength expected 
for the \ion{O}{vii}
absorption line 
($\sim 21.6$ \AA)
at redshift zero. 
Between
23 \AA\ and 24 \AA,
where redshifted
\ion{O}{vii} transitions
are expected,
the residuals are systematically
positive.  This structure
is a candidate broad emission
line.
Absorption from other transitions
of ionized oxygen
are also expected
in this crowded spectral region
\citep[e.g.,][]{det2011}.\\
We repeated this exercise
for the HETGS spectrum. 
Guided by the knowledge
of the RGS spectrum,
we were able to recognize 
 in the MEG the
same absorption trough
at $\sim$ 18 \AA.
In the MEG this falls towards 
the end of the sensitive
band, where the effective area
starts degrading. 
Blueward of this,
between 10 and 15 \AA,
the HETGS residuals show
some candidate absorption lines 
from the main \ion{Ne}{ix}--\ion{Ne}{x}
transitions, indicating
that some photoionized absorption
may affect this spectrum.

\subsection{Setup of the joint RGS/HETGS fit.}
\begin{figure}[t]
    \includegraphics[angle=180,width=0.5\textwidth]{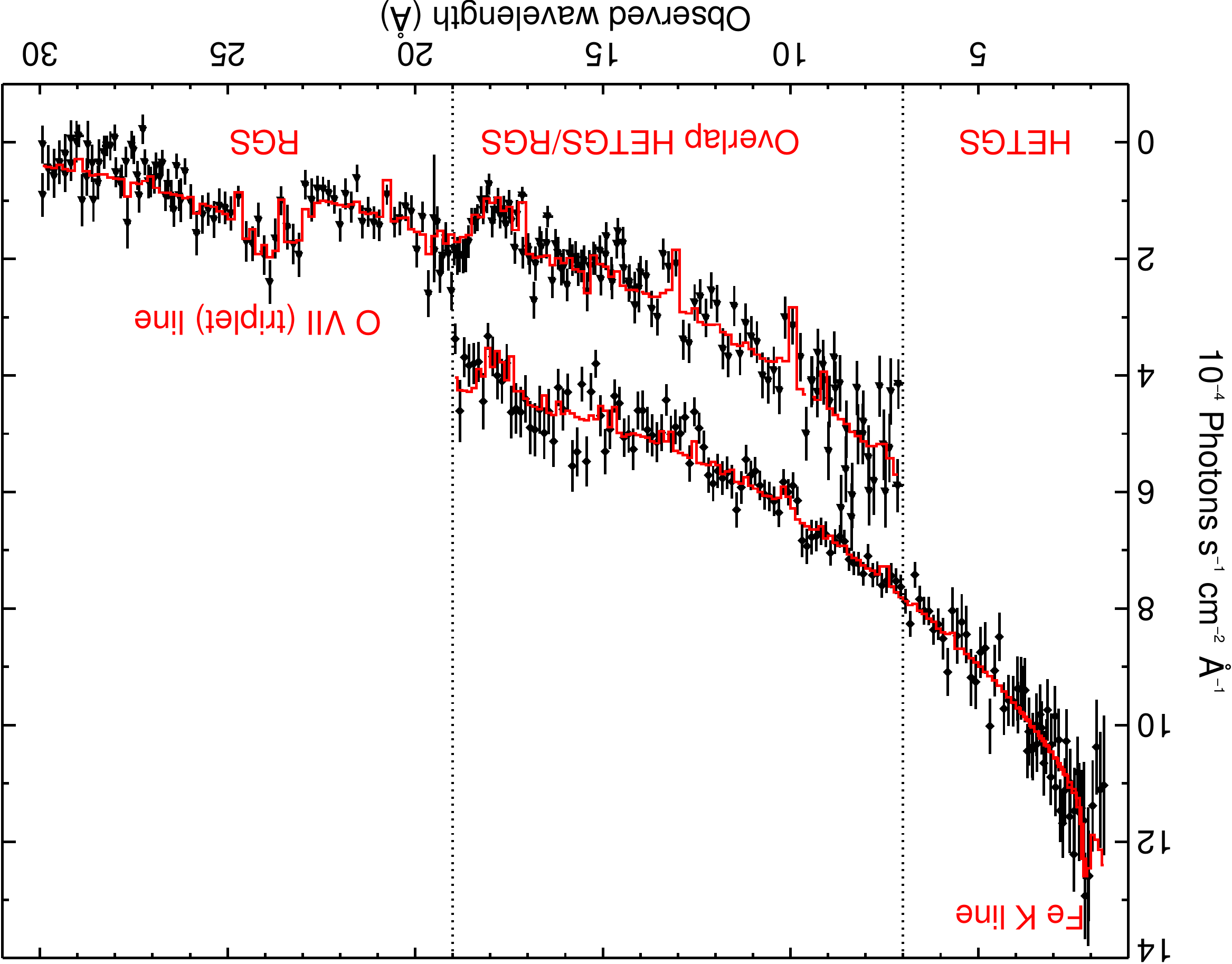}
  \caption{Best fit of the
    RGS and of the HETGS spectrum of 4C +74.26. The diamonds and the triangles
   indicate the HETGS and the RGS data points, respectively. The HETGS spectrum
   has been shifted upwards ($\times$ 3) for display purpose.
    Vertical lines indicate the band where the instruments overlap. 
    The solid lines represent our best fit models.
   Emission lines are labeled.
   The spectra have been rebinned for clarity.
  }
  \label{spec_qc.fig}
\end{figure}

The qualitative analysis of 
the RGS and  MEG
spectra shows hints of a complex
ionized absorption in this source.\\
In order to accurately disentangle
the multiple absorption components
of this spectrum,
we fit jointly
the RGS and the HETGS datasets.
The negligible variation
in observed flux during the
$\sim$4 months separating
these two observations 
(Table \ref{obs5.tab})
is indeed an indication
that the source and the absorbers 
were in the same conditions
when these two spectra were taken.
With a joint fit we take
advantage of  the high
HETGS spectral resolution at short
wavelengths, where most of the
features from higher ionization 
species are expected, 
and of the high sensitivity of the RGS at
long wavelengths, where
the absorption features
of e.g., ionized iron
and oxygen reside.\\
In the fit, 
we used the RGS between
7 and 30 \AA\ and 
the MEG between 
2 and 19
\AA. The quality
of the HEG spectrum 
is worse than the MEG, thus
we use HEG only in the
\fek \th region between
1.5 and 5 \AA.
For the joint fit,
we created two spectral
sectors in SPEX, 
one for the RGS
(RGS 1 and RGS 2) 
and one for the HETGS 
(HEG and MEG). 
In this way,
each instrument 
is fitted independently, 
but the model parameters
can be coupled.
In the following,
we fit
jointly the HETGS and the RGS
tying the absorption 
components together
but allowing
the continua
to vary.
In Fig. \ref{spec_qc.fig}
we show the final
best fit model in
the total energy 
range covered.

\begin{figure}[t]
    \includegraphics[angle=180,width=0.5\textwidth]{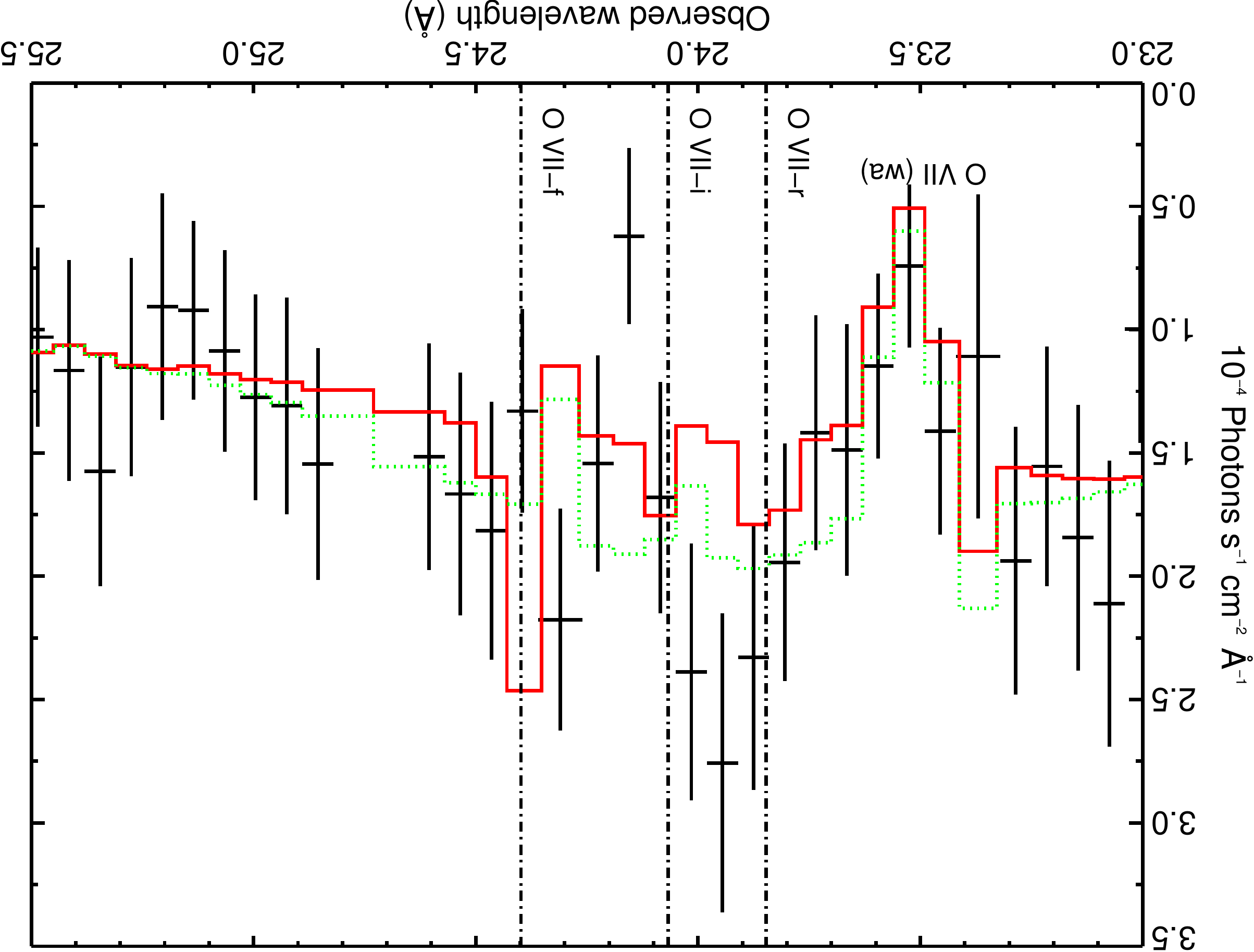}
  \caption{RGS-1 spectrum of 4C +74.26 in the 23.0--25.5 wavelength region. 
  The vertical dash-dotted lines mark the positions of the resonance, intercombination,
  and forbidden lines. 
  The solid line represents the fit with a narrow profiled \ion{O}{vii} triplet,
   which we were forced to reject (see Sect. \ref{em.ss}).
  The dotted line represents our best fit model with a single Gaussian line
  representing a broadened \ion{O}{vii} triplet.
  In both cases, the triplet profile is affected by absorption from
  ionized oxygen (\ion{O}{vii}) 
  intrinsic to the source
  as labeled. The spectrum has been rebinned for clarity.}
  \label{o7_qc.fig}
\end{figure}


\begin{figure}[!h]
 \centering
  \begin{minipage}[t]{0.44\textwidth}
    \includegraphics[width=1.0\textwidth,angle=180]{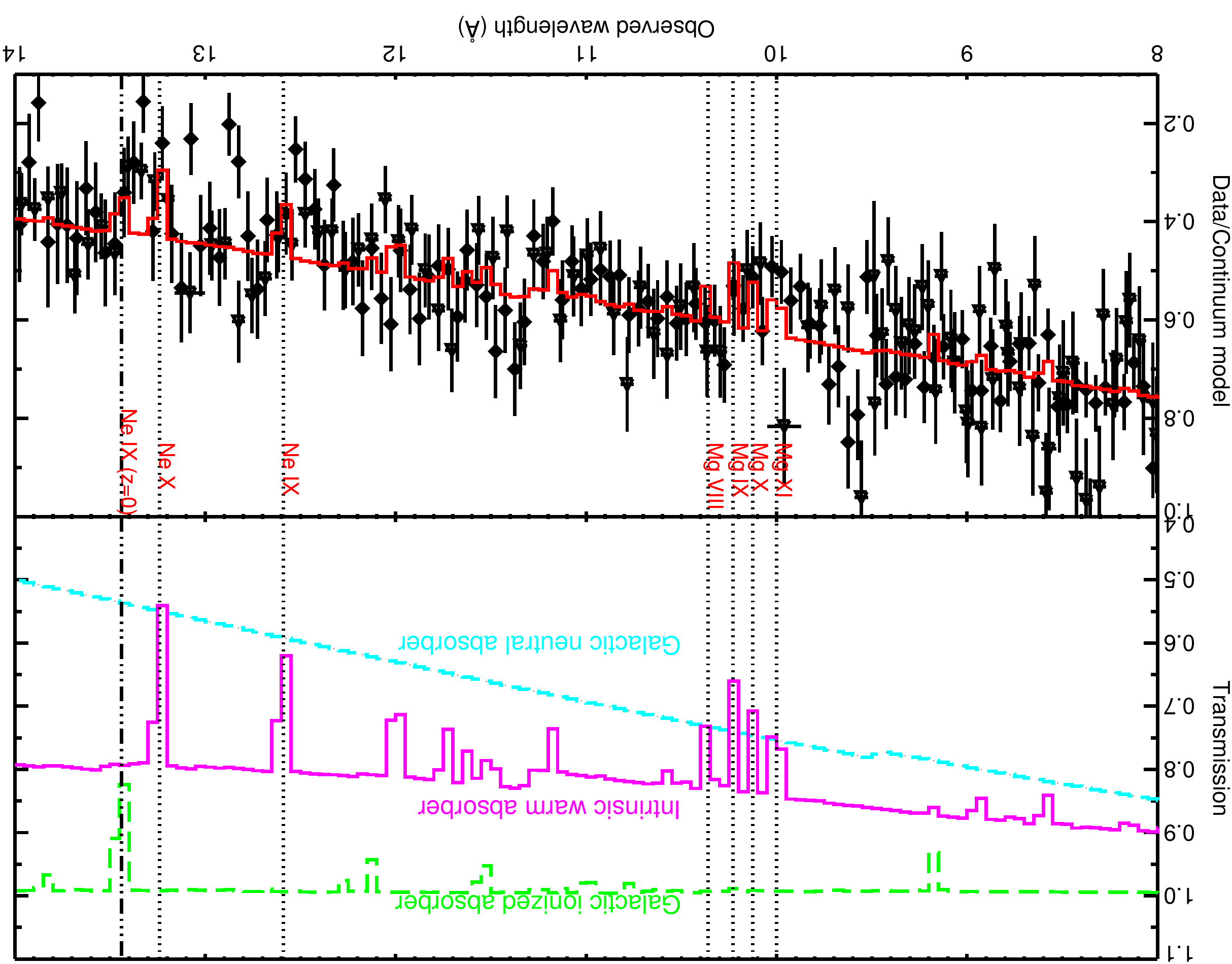}
     \vspace{0.5 cm}   
    \includegraphics[width=1.0\textwidth,angle=180]{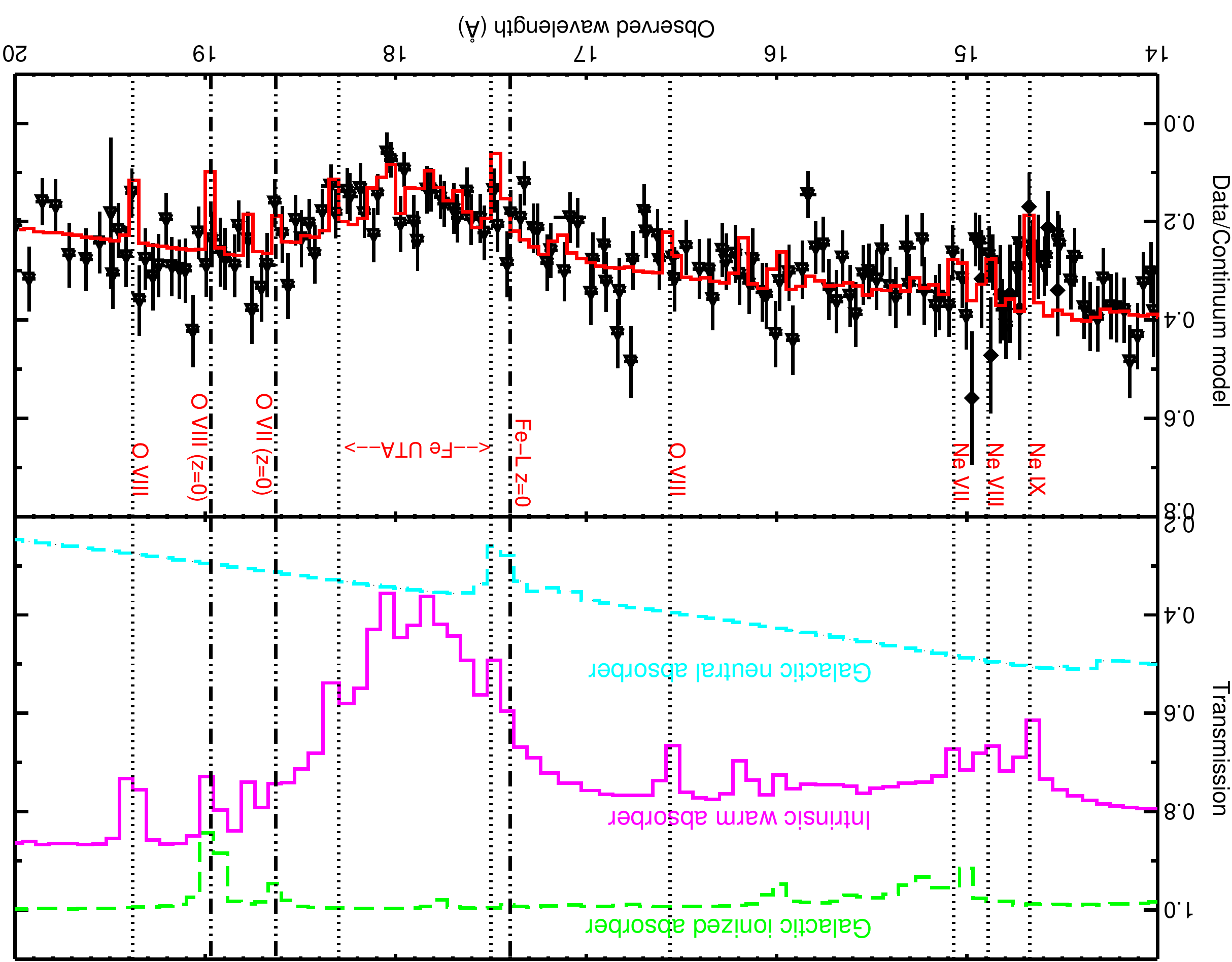}
    \vspace{0.5 cm}
    \includegraphics[width=1.0\textwidth, angle=180]{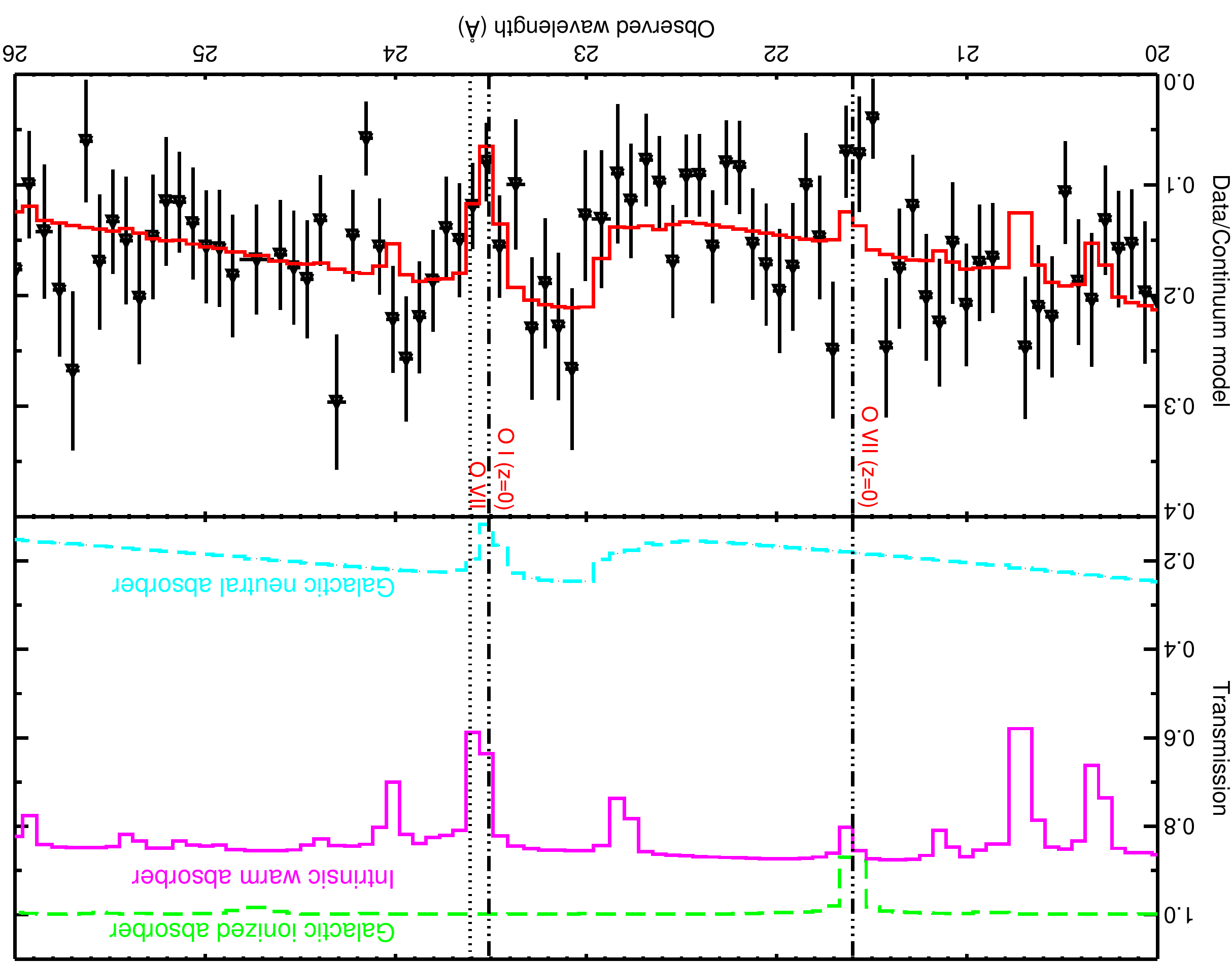}
  \end{minipage}
  \caption{From  top to  bottom: Absorption features of 4C +74.26
  in the 8--14 \AA\ (\textit{upper figure}), 14--20 \AA\
  (\textit{middle figure}), and 20--26 \AA\ band (\textit{lower figure}).
  In each figure the transmission (\textit{upper panel})
  of the Galactic (long-dashed lines) and intrinsic (solid line) absorbing components
  is shown together with the ratio between the data and the continuum
  model (\textit{lower panel}), highlighting the absorption features.
  The HETGS and the RGS data points are plotted as diamonds and triangles, respectively. 
  The solid line represent the best fit model. Vertical lines mark the position
  of the most prominent Galactic (double-dot-dashed lines) 
  and intrinsic (dotted lines) absorption features.}
\label{absqc.fig}
\end{figure}

\subsection{Continuum}
\label{contqc.s}
We set a simple continuum model
comprising a power law and
a phenomenological modified blackbody
mimicking a soft-excess 
\citep{sin1985} component.
For the latter we used the MBB 
model in SPEX, which includes
the effect of Compton scattering
\citep{kaa1989}.
In the fit we always kept
the modified blackbody
temperature of the HETGS
model  coupled to the RGS value
because in our fit
the band where
the soft-excess component
is supposed to dominate
is mostly covered by the RGS.
For the power-law component, 
we set the initial value 
of the slope to that determined 
by the EPIC-pn fit. As the best fit was reached, 
the value of $\Gamma$ settled 
at $1.67\pm0.05$, while
we found $T_{\rm BB}=170 \pm 15$ eV.
The final best fit
values for the continua
are given in Table \ref{parqc.tab},
first and second panel. 
A small difference in the normalizations
of the continuum components
is sufficient to account
for the change of flux
between the HETGS 
and the RGS observations.

\begin{table*}[t]
\caption{Best fit parameters and errors for the final best fit model.
Parameters without errors were kept coupled in the fit
either to the RGS or to the HETGS value.}     
\label{parqc.tab}      
\centering                    
\begin{tabular}{lcccc}        
\hline\hline                 
Model component & Parameter & RGS value & HETGS value & Units \\
\hline
\multirow{2}{*}{Power law} & 
$\Gamma $ \tablefootmark{a}& $1.67 \pm 0.05$ &$ 1.68 \pm 0.02$ & \\
& $L_{\rm 0.3-10.0 \, keV}$\tablefootmark{b} & $13.7 \pm 0.5$ & $11.4 \pm 0.3$ & $10^{44}$ \ergs \\
\hline
\multirow{2}{*}{Blackbody} 
& $T_{\rm BB}$ \tablefootmark{c}& $170 \pm 15$ & 170 & eV \\
&$L_{\rm 0.3-10.0 \, keV}$ \tablefootmark{b}& $1.7 \pm 0.8$ & $1.0 \pm 0.8$ & $10^{44}$ \ergs \\
\hline
\multirow{3}{*}{\fek \th line} 
& w \tablefootmark{d}& 1.92 & $1.92 \pm 0.08$ & \AA \\
& \fwhm \tablefootmark{e}& 0.16 & $0.16\pm0.13$ & \AA \\
& $L_{\rm Fe \, K\alpha \,line}$ \tablefootmark{b}& 7 & $6 \pm 4$ & $10^{42}$\ergs \\
\hline
\multirow{3}{*}{\ion{O}{vii} (triplet) line} 
& w \tablefootmark{d} & $21.7 \pm 0.1$ & 21.7 & \AA \\
& \fwhm \tablefootmark{e} & $ 1.0 \pm 0.3$ & 1.0 & \AA \\
& $L_{\rm \ion{O}{vii} \,line}$\tablefootmark{b}& $9 \pm 3$ & 9 & $10^{42}$\ergs \\
\hline
\multirow{3}{*}{Warm Galactic absorber} 
& \nh \tablefootmark{f}& $ 2 \pm 1$ & 1.2 & $10^{20}$ \colc \\
& $T$\tablefootmark{g} & $0.23 \pm 0.04$& 0.20 & keV \\
\hline
\multirow{3}{*}{Intrinsic warm absorber} 
& \nh \tablefootmark{f}& $3.5 \pm 0.6 $ & $3.1 \pm 0.3 $ & $10^{21}$ \colc \\
& $\log \xi$ \tablefootmark{h}& $2.5 \pm 0.1$ & $2.57 \pm 0.06$ & \xic\\
& $v_{\rm out}$\tablefootmark{i} & 3600 & $3600 \pm 70$ & \kms \\
\hline
\end{tabular}
\tablefoot{
\tablefoottext{a}{Power-law slope.}
\tablefoottext{b}{Model component luminosity, in the quoted band.}
\tablefoottext{c}{Blackbody temperature.}
\tablefoottext{d}{Wavelength of the line centroid.}
\tablefoottext{e}{Full width at half maximum of the line.}
\tablefoottext{f}{Absorber column density.}
\tablefoottext{g}{Absorber temperature.}
\tablefoottext{h}{Absorber ionization parameter.}
\tablefoottext{i}{Absorber outflow velocity.}
}
\end{table*}


\begin{table*}
\caption{Identification of the main absorption features in the RGS and in the HETGS spectrum of 4C +74.26.}     
\label{lines.tab}      
\centering                    
\begin{tabular}{lccccc}        
\hline\hline                 
Wavelength \tablefootmark{a} &
 z \tablefootmark{b} &   
 $v_{\rm out}$ \tablefootmark{c} & 
 $\Delta C$\tablefootmark{d} 
  & \multicolumn{2}{c}{Identification}\\
\AA & &\kms & & Ion & Transition\\
\hline
21.602 & 0.104 & $2100\pm900$ & -9 & \ion{O}{vii} & $1s^2-1s2p^1 P_1$ \\
21.602 & 0 & 0 & -14 & \ion{O}{vii} &  $1s^2-1s2p^1 P_1$\\
13.447 & 0.104 & $3900\pm200$ &-96 & \ion{Ne}{ix} & $1s^2-1s2p^1 P_1$\\
13.447 & 0 & 0 & -47 & \ion{Ne}{ix} & $1s^2-1s2p^1 P_1$\\
12.132 & 0.104 & $4100\pm300$ &-52& \ion{Ne}{x} & $1s-2p $ (\lia)\\
\,9.378 \tablefootmark{*} & 0.104 &   \multirow{3}{*}{$3600 \pm 100$} & \multirow{3}{*}{-85} &\ion{Mg}{ix} & $2s^2-1s2s^22p$\\
\,9.281\tablefootmark{*} & 0.104 & & & \ion{Mg}{x} & $1s^2(^1S)2s-1s(^2S)2s2p(^3P^0)$\\
\,9.169 \tablefootmark{*} & 0.104 & & & \ion{Mg}{xi} & $1s^2-1s2p^1P_1$\\
\hline
\end{tabular}
\tablefoot{
\tablefoottext{a}{Nominal laboratory wavelength of the line.}
\tablefoottext{b}{Redshift applied.}
\tablefoottext{c}{Blueshift applied.}
\tablefoottext{d}{Improvement of the C-statistics with respect to a model including only the continuum, the emission lines, and the
Galactic neutral absorber.}
\tablefoottext{*}{Lines fitted simultaneously with the same blueshift.}}
\end{table*}


\subsection{Emission lines}
\label{em.ss}

The presence of a broad \fek \th
emission line in \qc is well
established \citep{bal2005,lar2008}.
The line is also clearly visible
in the HETGS data (Fig. \ref{spec_qc.fig}).
We fitted it with a phenomenological
Gaussian emission line with free centroid,
width, and normalization. 
The values we obtained
(Table \ref{parqc.tab}, third panel)
are  consistent both with what
is reported in the literature \citep{bal2005a} 
and with our phenomenological
fit of the EPIC-pn (Sect. \ref{sedqc}).\\
In Fig. \ref{o7_qc.fig}, 
we show the
spectrum in the 23--25.5 \AA\ range
where we already noticed an
excess in the residuals (Sect. \ref{resqc})
reminiscent of a broad emission line.
At first, we tested whether these
residuals could be accounted for
with a narrow-profiled \ion{O}{vii}
triplet. We added to the fit three
delta-profiled emission lines
(DELT model in SPEX) for the
resonance ($\lambda$=21.6 \AA), 
intercombination ($\lambda$=21.8 \AA),
and forbidden line ($\lambda$=21.1 \AA).
We left the normalization
of the forbidden line free to vary
and we assumed a ratio 1:3 for the
other lines, as expected 
if photoionization occurs
in a low-density plasma
approximation \citep{por2000}. This
fit (Fig. \ref{o7_qc.fig})
does not reproduce
 the data well and leaves
large residuals between 23
and 24 \AA.\\
Thus, we added to the fit
a Gaussian profiled 
emission line. 
We left the line centroid free to vary
among the nominal wavelengths of the
\ion{O}{vii} triplet and we
used the width of the broad 
$\rm H_{\rm \alpha}$ line given
in \citet{win2010} to set
the fitting range
for the width of
a blended triplet
(\fwhm=[0.36--1.23]\AA).
A broad \ion{O}{vii} line
having \fwhm=$1.0\pm 0.3$
better accounts for the excess in the
residuals in the 23--25.5 \AA\
region. 
For the final
fit, the statistical improvement
produced by the addition of
the \ion{O}{vii} broad line
is $\Delta C=-24$ for 
three additional degrees
of freedom. An F-test gives
a probability of a chance
improvement of $\sim 10^{-5}$.
In this fit the normalization
of the narrow components
goes to zero,
indicating that the data quality
does not allow them  to be deblended
 from the broad component.
The modeling of the broad
emission line is critical
for a correct evaluation
of the absorption 
\citep[e.g., ][]{cos2007,dig2013}
because many transitions from
ionized oxygen may in principle
be detected within the line 
profile.
We outline in Table \ref{parqc.tab},
third and fourth panel, 
the final best fit values
for the line parameters. 

\subsection{Line-by-line fitting of the absorption features}
\label{linefit.s}

Before proceeding with a global modeling
of the absorbing components, we first attempted
to identify the absorption features of the
spectrum on a line-by-line basis
\citep[e.g.,][]{ebr2013}.
We note, however, that not all
the WA features can be identified
with this method because
of blending with neighboring transitions
(e.g., the Fe-UTA)
or with other components (e.g., Galactic).
Moreover, only a global modeling
is able to account  for the additional continuum
curvature produced, for example,  by  an ionized
absorber.\\
We visually identified in the spectrum
the most prominent features, and
for each of these 
we added to the model a Gaussian 
profiled absorption line multiplied
by a blueshift model.
The line centroid was set  to the wavelength 
of the nearest known transition, while the line FWHM
was set to the default value of 0.1 \AA. Thus,
in this exercise, the free parameters
were the line normalization and the
blueshift. \\
In Table \ref{lines.tab} we list our line
identifications. In the RGS band
we detected an \ion{O}{vii} resonance line
at redshift zero and at the redshift of
4C +74.26. The addition of
an \ion{O}{viii}-\lia\ line ($\lambda=18.97$ \AA)  line
at the redshift of the source
resulted instead in a negligible
improvement of the fit ($\Delta C$=-3).
 In the HETGS band, we detected
absorption from \ion{Ne}{ix} at redshift 0
and from  \ion{Ne}{ix}, \ion{Ne}{x},
\ion{Mg}{ix},  \ion{Mg}{x}, and \ion{Mg}{xi} 
in the source rest-frame. The magnesium
lines are blended, so we fitted them
simultaneously with the same
blueshift. All
the lines detected in 
the HETGS band show a
similar blueshift, suggesting
that they may be part
of the same outflowing system.
\subsection{Absorption at redshift zero}
\label{galabs}
We modeled the Galactic cold
absorption using a collisionally
ionized plasma model in SPEX
(HOT), setting a temperature
of 0.5 eV for the neutral gas
case. This component produces
\ion{O}{i} and \ion{Fe}{i}
absorption at $\sim$23.5 \AA\ and
 $\sim$17.4 \AA, respectively.\\ 
As pointed out in Sect.
\ref{resqc} narrow absorption lines
from \ion{O}{vii} and \ion{Ne}{ix}
at redshift zero are detected  respectively
in the RGS and in the HETGS spectrum.
These could originate in the warm
plasma of the Galactic corona,
which is collisionally ionized
\citep[e.g.,][]{yao2005,pin2012}.
To model it, we added another
HOT component to the fit.
We left both $T$ and 
the gas column density 
\nh free to vary. We kept instead
the broadening velocity frozen
to the default value of 100 \kms.
The final best fit values
that we found for all these 
free parameters are listed
in Table \ref{parqc.tab}, 
fifth panel.

\subsection{Intrinsic photoionized absorption}
\label{wa}

We modeled  the intrinsic 
photoionized absorption 
using the XABS model 
in SPEX which computes the transmission 
of a slab of material 
where all the ionic column densities are linked 
to each other
through the photoionization balance
prescribed by the SED
(Fig. \ref{sedqc.fig}).
We computed the SED 
with the SPEX auxiliary tool XABSINPUT
and the photoionization code Cloudy \citep{fer2013}, 
version 13.01. 
For the XABS component, 
we allowed the column density, the
ionization parameter, and
the outflow velocity of the gas to
vary, while we kept the broadening 
velocity frozen to the default value
of 100 \kms.\\
We found that an intrinsic photoionized
absorber with 
$\nh \sim 3 \times 10^{21}$ \colc\
and $\log \xi \sim 2.6$
best fits
the candidate absorption features 
of the spectrum. The systematic blueshift 
of the lines corresponds 
to an outflow velocity of
$v_{\rm out}\sim 3600$ \kms.
We list in Table \ref{parqc.tab}, sixth panel,
the best fit parameters for the WA.
After achieving the best fit, 
we decoupled the column density
and the ionization parameter
of the RGS model from the HETGS values
to check for a possible
time variability of the WA
in the 4 months 
separating the HETGS from
the RGS observation.
We found that  during this time interval 
the WA parameters are consistent
not to have varied.
We note that a
one-zone WA is sufficient
to best fit the ionized
absorption features of
the spectrum. Indeed,
the fit erases any additional
ionized absorbing
components, either
photoionized or collissionally
ionized.
The final C-statistics for
a model including two
Galactic absorbers and
an intrinsic WA
is C/Expected C=1065/931.\\
In Fig. \ref{absqc.fig}, 
we show the transmission of all
the absorbing components 
of the model, together
with the ratio between
the data and the continuum
model, which highlights
the absorption features.
In the RGS band the most
evident WA feature
is the broad absorption
trough visible at $\sim$ 18 \AA.
This
is mostly produced by 
the unresolved transitions array (UTA) 
from the ionized iron 
(e.g., \ion{Fe}{x}--\ion{Fe}{xx})
contained in the photoionized
gas. In addition,
a \ion{O}{vii} absorption line
is prominent at $\sim$23.5 \AA.
This feature is blended with
the \ion{O}{i} line from
the neutral absorber in
the Galaxy. In the HETGS band,
the absorption lines are weak.
The most apparent features
are from highly ionized
species, such as
\ion{Ne}{vii}--\ion{Ne}{x}
and \ion{Mg}{viii}--\ion{Mg}{xi}.


\section{Discussion}
\label{dsou}
%
%
%

%
%
%
%
%
%
We have presented 
a joint analysis of the RGS 
and   HETGS spectra
of the heavily X-ray
absorbed radio-loud quasar
4C +74.26.
Thanks to the high spectral
resolution of these grating spectra,
we were able to reveal a rich
spectrum of absorption
features originating from
both Galactic
and intrinsic material. 
In our analysis
we used the total Galactic
column density given
in \citet{wilh22013}, 
which includes the contribution
of molecular hydrogen.
This is roughly twice
the value provided
by 21 cm surveys.
The enhanced Galactic
absorption explains
the heavy suppression
of the soft X-ray
flux that was noticed
in the past for this
source \citep{bri1998,
sam1999,ree2000,has2002,
bal2005}.\\
The intrinsic
absorption comprises 
a highly ionized WA
which produces 
a deep Fe-UTA trough in the RGS
and the weak absorption
features that are visible in the
HETGS spectrum. We found that
an outflow velocity of $\sim 3600$ \kms\
is required for a  best fit of the absorption
features visible in the two spectra.
This finding is
  evidence for WA absorption 
in radio-loud objects, which so far
has been scarce.
Indeed, in addition to
3C 382, 3C 445 and 3C 390.3,
\qc is the fourth radio-loud source
where a photoionized outflow 
has been clearly characterized 
in a high-resolution 
dataset.
The column density, ionization
parameter, and outflow
velocity that we measured for
the WA in 4C +74.26 are
within the
range observed in Seyfert 1
galaxies \citep{mck2007}
and are also in line with the values
found in 3C 382 and
3C 390.3, the other
two radio-loud galaxies
hosting a classical
WA. The case of 3C 445
is an outlier, as this source
hosts a high-velocity, 
high-column UFO-like wind
\citep[see the review of][]{tor2012}.\\
In the following sections
we use the results of our
spectral analysis and the
information from the literature
to infer a possible geometrical
model for the outflow is this AGN.
To this purpose,
in Sect. \ref{dion}
we estimate the possible location 
and the energetics of
the warm absorber.
In Table \ref{disc.tab},
upper panel, we outline
some basic physical
properties of the source that
serve for an order of magnitude
comparison.
We took the black hole
mass $M_{\rm BH}$
and the source inclination $i$
from the literature
as already explained 
in Sect. \ref{intro5}.
From a numerical integration
of the SED of Fig. \ref{sedqc.fig}
we computed the ionizing luminosity
$L_{\rm ION}$ between 1 and 1000 Ry
and the bolometric luminosity
over the whole optical and
X-ray band. We note that
the bolometric luminosity is
probably underestimated
because the radio emission at
low energies and the gamma ray emission
at high energies are not included
in our SED.
Hence, using these data
we estimated the Eddington
luminosity $L_{\rm Edd}$
and the mass accretion rate $\dot{M}_{\rm acc}$, 
for which we assumed an accretion efficiency
$\eta=0.1$. 
For the jet power $P_{\rm jet}$ 
we used the radio flux
at 1.4 GHz \citep{con1998} 
and the scaling relationship 
of \citet{cav2010}. 
The radius of the 
broad line region 
$R_{\rm BLR}$
scales with the optical luminosity
at 5100 \AA\ \citep{wan2002}.
The luminosity value  is given in
\citet{win2010}.
Finally, 
the radius of the putative torus $R_{\rm TOR}$,
which is nominally set by the dust sublimation radius,
scales with $L_{\rm ion}$
\citep{kro2001}.

\subsection{Location and energetics of the ionized outflow}
\label{dion}
\begin{figure}[t]
    \includegraphics[width=0.5\textwidth]{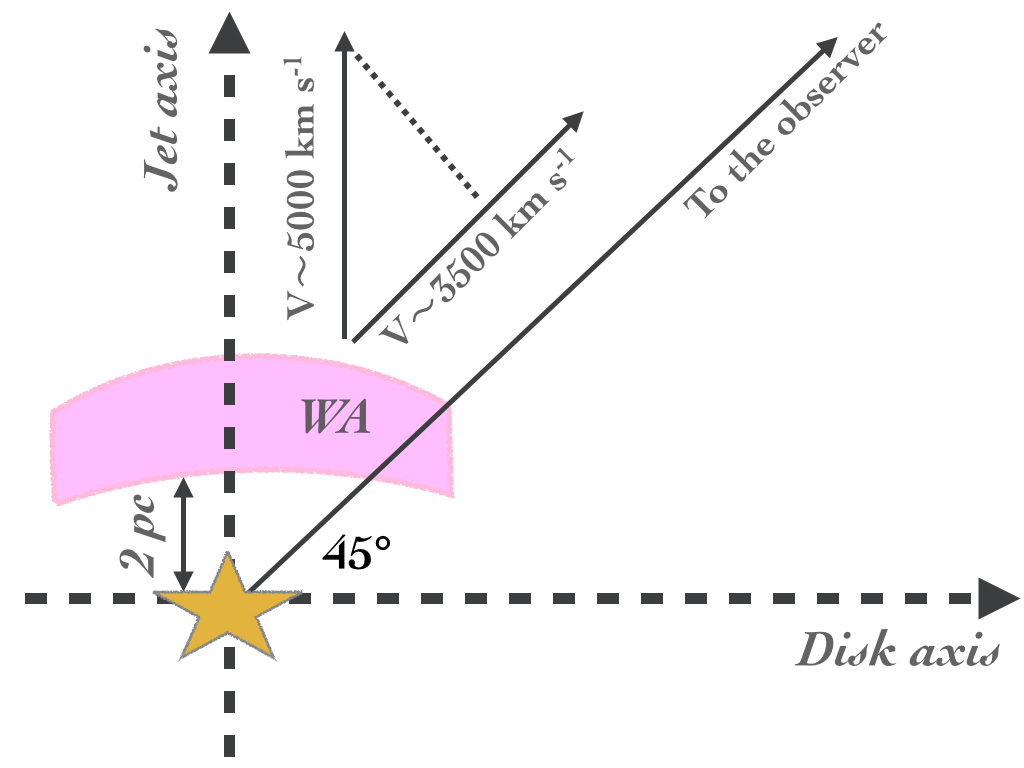}
  \caption{Outflow in the inner region of 4C +74.26.
 The observer's line of sight lies at 45$^{\circ}$ from the jet axis.
The WA is part of a polar outflow located outside the BLR.
The ionized gas  outflows along the polar direction with a velocity
of $\sim$5000 \kms, which is observed as $\sim$3500 \kms\
from the observer's inclination angle.}
\label{geo_qc.fig}
\end{figure}

In Table \ref{disc.tab}, 
lower panel, we outline
some physical properties 
of the ionized
outflow that we estimated
using our measured parameters,
namely
$\nh\sim 3.1\times 10^{21}$ \colc,
$\log \xi \sim 2.6$, and
$v_{\rm out}\sim 3600$ \kms.
We follow here the argumentation
of  \citet{blu2005}, which
assumes that the outflow 
is a partially filled
spherical shell of gas with 
a volume filling factor $f$.
An analytical expression for the volume
filling factor $f$ is derived 
in \citet{blu2005} from
the prescription that  the kinetic
momentum of the outflow 
must be on the order of the momentum 
of the absorbed radiation plus the momentum 
of the scattered radiation. 
For the ionized outflow in \qc 
we found that the
ionized gas fills only
 $\sim 0.007 \%$ of the
spherical volume, which suggests that
it may consist of sparse clumps.\\
We set a range of possible distances
for the absorber from the conditions
that the velocity of the outflow 
must exceed the escape velocity from the
AGN and that the outflowing shell must not 
be thicker than its distance  
from the center ($\Delta r / R \leq 1$).
Analytically,
$$
\frac{2GM_{\rm BH}}{v^2}
\leq
 R
 \leq
\frac{L_{\rm ion} f}{\xi \nh},$$
where G is the gravitational constant.
For our parameters,
both these expressions return
a value of $\sim$2 pc (Table \ref{disc.tab}).
This constrains
the ionized outflow of \qc to be located
outside the BLR 
($R_{\rm BLR}=0.2$ pc),
but within the boundary
of the putative torus 
($R_{\rm TOR}=6$ pc). \\
A patchy ionized outflow located
outside the BLR is a natural
candidate for being
the scattering outflow that is
required in the \citet{rob1999} 
analysis of the polarized
optical spectrum of this
source. Their model prescribes
that the observed redshift
of the polarized H$\alpha$ 
line is due to a high-velocity
motion of the scattering material
which polarizes the BLR light.
In this framework, the
outflow velocity inferred
for the scatterer
depends on the inclination
of the scattering cone 
with respect to the jet axis.
For the case of a scattering outflow
coaligned with the radio jet, they
quote a velocity of $\sim$5000 \kms.
Interestingly, if we consider
the same source inclination 
used in the \citet{rob1999}
model ($\sim 45^{\circ}$)
and we assume that the WA
found in our analysis
is outflowing along
the polar axis of the source,
we obtain a deprojected
velocity of $v_{\rm out}/\cos 45^
{\circ}\sim$5000 \kms\
(Fig. \ref{geo_qc.fig}).
This matches the 
\citet{rob1999} prediction.
This correspondence hints
at the possibility that 
the WA detected here
and the outflowing polar scatterer 
discovered in \citet{rob1999} are one 
and the same.\\
Given the velocity,
the mass outflow rate is given by
$$
\dot{M}_{\rm out}=
\frac{1.23 m_{\rm p} L_{\rm ion} f v_{out} \Omega}{\xi},
$$
where $m_p$ is the proton mass
and $\Omega$ is the solid
angle of the outflow, which
we set to 2.1 sr, 
as in \citet{tor2012}.
This is derived assuming that
at least 50\% 
of radio-loud objects host
an outflow, like in a Seyferts galaxy,
and using the information
that $\sim$33\% 
of the radio galaxies belonging to
the 3CR sample are type 1 AGN
\citep{but2009}. Hence, using the mass
outflow rate, the kinetic
luminosity of the outflow is 
readily computed as 
$L_{\rm kin}=\frac{1}{2}
\dot{M}_{\rm out}
v_{\rm out}^2 $.\\
The value we obtained
for the kinetic luminosity is
at least four 
orders of magnitude lower
than the bolometric luminosity.
Theoretical AGN feedback models
\citep[e.g.,][]{dim2005, 
hop2010} typically require
kinetic luminosities
comparable with the
bolometric luminosity
for an outflow
to be able to halt
the star formation in
a typical galactic bulge.
Thus, this outflow is
unable to deliver
a significant feedback
in this AGN.
Moreover,
as found for the
other radio-loud galaxies
hosting a WA,
the kinetic luminosity
of the outflow is
negligible compared
to the jet power (
$L_{\rm kin} \sim
10^{-2} P_{\rm jet}$).
Thus, the case of
4C +74.26 confirms that
the jet is a more likely
driver of AGN feedback
in radio-loud galaxies
\citep{tor2012}. \\

\begin{table}
\caption{Properties of 4C +74.26.}     
\label{disc.tab}      
\centering                    
\begin{tabular}{lc}        
\hline\hline 
Source properties & Ref \\
\hline
$M_{\rm BH}= 3 \times 10^{9}$ \msun  & \citet{win2010} \\
$i \leq 49^{\circ}$ & \citet{pea1992}\\
$L_{\rm bol}=9.7 \times 10^{46}$ \ergs & Sect. \ref{dsou} \\
$L_{\rm bol}/L_{\rm Edd}=0.25$ &Sect. \ref{dsou} \\
$\dot{M}_{\rm acc}=17$ \msunyr & Sect. \ref{dsou}\\
$L_{\rm ion}=8.8 \times 10^{46}$ \ergs & Sect. \ref{dsou}\\
$P_{\rm jet}=2 \times 10^{44}$ \ergs &Sect. \ref{dsou} \\
$P_{\rm jet}/L_{\rm Edd}=6 \times 10^{-4}$ & Sect. \ref{dsou}\\
$R_{\rm BLR}=0.2$ pc & Sect. \ref{dsou}\\
$R_{\rm TOR}=6$ pc & Sect. \ref{dsou}\\
\hline 
Ionized outflow properties & Ref\\
\hline
1.6 $ \leq R \leq 1.8$ pc&  Sect. \ref{dion}\\
$f=7 \times 10^{-5}$ &  Sect. \ref{dion}\\
$\dot{M}_{\rm out}=0.4$ \msunyr &  Sect. \ref{dion}\\
$L_{\rm kin}=1.5 \times 10^{42}$ & Sect. \ref{dion}\\
\hline
 \end{tabular}
 \end{table}
%
%

%
\section{Summary}
\label{concqc}

We performed a joint analysis
of the RGS and  HETGS spectra
of the radio-loud quasar 4C +74.26.
The spectrum is
affected by a heavy X-ray
absorption arising
from both Galactic
and intrinsic material.\\ \\
 Most of the
absorption in the soft X-ray band
is due to the Galactic
ISM. We point out
that when also considering 
the contribution of molecular hydrogen,
the total Galactic \nh is
roughly twice the
standard value provided 
by 21 cm surveys.\\ \\
A photoionized outflow 
($\nh \sim 3.2 \sim 10^{21}$ \colc,
$\log\xi \sim 2.6$,
$v_{\rm out} \sim 3600$  \kms)
located at the source rest-frame
produces a sharp Fe-UTA trough in the RGS 
and the weak absorption
features visible in the HETGS.
The kinetic luminosity
carried by the outflowing
gas ($L_{\rm kin} 
\sim 10^{-5}
 L_{\rm bol}$) is negligible for
the AGN feedback in this
source.\\ \\
We discuss a scenario where the photoionized gas is
part of a polar-scattering outflow, also detected 
in the optical-polarized spectrum.

%

\begin{acknowledgements}
The scientific results are based on 
data obtained from the Chandra and the  XMM-Newton data archives. 
SRON is supported financially by NWO,
the Netherlands Organization for Scientific Research.
LDG acknowledges support from the Swiss National Science
Foundation. We thank Enrico Piconcelli, Margherita Giustini, and
Francesco Tombesi for useful discussions. We thank
Jelle Kaastra and Missagh Mehdipour
for commenting and carefully reading this manuscript.

\end{acknowledgements}

\end{document}